\documentclass[prb,twocolumn]{revtex4}

\usepackage{graphicx}
\usepackage{amsmath}
\usepackage{amsbsy}
\usepackage{bm}

\begin{document}

\newcommand{\notop}{{{}}}
\newcommand{\Egy}{E}
\newcommand{\ksa}{k\sigma\ell}
\newcommand{\cdag}{c^\dag}
\newcommand{\cno}{c^{{}}}
\newcommand{\up}{\uparrow}
\newcommand{\down}{\downarrow}
\renewcommand{\d}{{\rm d}}
\newcommand{\e}{{\rm e}}

\newcommand{\ket}[1]{|#1\rangle}
\newcommand{\bra}[1]{\langle #1|}
\newcommand{\ketbra}[2]{| #1 \rangle\!\langle #2|}
\newcommand{\dens}[1]{\rho_{#1}^{{}}}
\newcommand{\ave}[1]{\left\langle #1\right\rangle}
\newcommand{\imai}{{\rm i}}
\newcommand{\eqlab}[1]{\label{eq:#1}}
\newcommand{\eqreff}[1]{Eq.~(\ref{eq:#1})}
\newcommand{\eqsref}[2]{Eqs.~(\ref{eq:#1},\ref{eq:#2})}
\newcommand{\iop}{{\rm i}0^+}
\newcommand{\Ed}{E_1^{{}}}


\title{Tunneling through nanosystems: Combining broadening with 
many-particle states}

\author{Jonas Nyvold Pedersen and Andreas Wacker}

\affiliation{Department of Physics, University of Lund, Box 118,
22100 Lund, Sweden.}  \date{\today}

\begin{abstract}
We suggest a new approach for transport through finite systems based
on the Liouville equation. By working in a basis of many-particle
states for the finite system, Coulomb interactions are taken fully
into account and correlated transitions by up to two different contact states
are included. This latter extends standard rate equation
models by including level-broadening effects.  The main result of the
paper is a general expression for the elements of the density matrix
of the finite size system, which can be applied whenever the
eigenstates and the couplings to the leads are known. The approach
works for arbitrary bias and for temperatures above the Kondo
temperature.  We apply the approach to
standard models and good agreement with other methods in their
respective regime of validity is found.
\end{abstract}
\pacs{73.23.Hk,73.63.-b} 

\maketitle

\section{Introduction}
Transport through nanosystems such as 
quantum dots and molecules has received 
enormous interest within the last
decade \cite{DattaBook1995,FerryBook1997,ImryBook2001}.
Typically this problem  is treated within one of 
two different approximations:
(i) Rate equations \cite{BeenakkerPRB1991} 
for electrons entering and leaving the system, which can
also take into account complex many-particle states in the central
region \cite{KinaretPRB1992,PfannkuchePRL1995}. Here broadening
effects of the levels are entirely neglected. It can be shown that
these rate equations become exact in the limit of high bias
\cite{GurvitzPRB1996}.
(ii) The transmission formalism, which is usually evaluated
by Green function techniques \cite{MeirPRL1992,HaugBook1996}
(alternatively,
scattering states can be calculated directly \cite{FrensleyRMP1990}),
allows for a consistent treatment of level broadening due to the
coupling to the contacts. In principle, many-particle effects
can be incorporated into this formalism, but
the determination of the appropriate self-energies is
a difficult task, where no general scheme has been found by now.
Thus, many-particle effects are usually considered on a mean-field
basis including exchange and correlation potentials 
\cite{diVentraPRB2002,BrandbygePRB2002,XueCP2002,HavuPRB2004}, which
are of particular importance for the transport through molecules. 
Mean-field calculations are well justified for extended systems, such
as double-barrier tunneling diodes \cite{PoetzJAP1989,LauxJAP2004},
which exhibit many degrees of freedom (e.g., in the plane
perpendicular to the transport). 
However, the bistability frequently obtained for such structures 
is questionable for systems
with very few degrees of freedom as studied here. 
See, e.g., the discussion in Sec.~III.B.4 of 
Ref.~\onlinecite{SprekelerPRB2004}.

In our paper we want to bridge the gap between these approaches
by considering the Liouville equation for the dynamics of the central
region coupled to the contacts. The approach works within a basis of
arbitrary many-particle states, thus fully taking into account
the interactions within the central region. While the first order
in the coupling reproduces previous work using rate equations
\cite{GurvitzPRB1998},
the second order consistently takes into account broadening effects.
This is analogous to the consistent treatment of broadening
for tunneling resonances in density-matrix theory.
 \cite{WackerAdvances2001}

The paper is organized as follows: We first present the formalism in
section \ref{SecFormalism}. Then we demonstrate its
application to the simple problem of tunneling through a single
level, section \ref{SecSingleLevel}. We show explicitly that the
exact Green function result is recovered for all biases and
temperatures. In sections \ref{SecDoubleDot} we give results
for the double-dot system with Coulomb interaction where both 
standard approaches fail. Finally we consider the spin-degenerate
single dot in section \ref{SecDoubleSpin}
to investigate Coulomb blockade as well as
the limit of low temperatures.

\section{Introducing the formalism}\label{SecFormalism}
The total Hamiltonian for the system consisting of leads and 
the dot can be written as
\begin{equation}
H=H_D+H_{\text{Leads}}+H_T.
\end{equation}
The first term describes the dot. Our key issue is the assumption that
the dot can be diagonalized in 
absence of coupling, and the (many-particle) eigenstates and
eigenenergies for $H_D$ are denoted $|a\rangle$ and $\Egy_a$. Thus we
have
\begin{equation}
H_{D}=\sum_{a}\Egy_a|a\rangle\langle a|.
\end{equation}
The leads are described by free-particle states 
\begin{equation}
H_{\mathrm{Leads}}=\sum_{\ksa}\Egy_{\ksa}\cdag_{\ksa} \cno_{\ksa}
\end{equation}
where $\sigma=\up,\down$ describes the spin, $k$ labels
the spatial wave functions of the contact states and  
$\ell$ denotes the lead. In the following
we assume two leads, i.e.  $\ell=L,R$, but generalization to
more leads is straightforward.
Finally, the last part in the Hamiltonian expresses the tunneling between
the states in the leads and the dot
\begin{equation}
H_T=\sum_{\ksa,ab}\left[T_{ba}(\ksa)\ketbra{b}{a} \cno_{\ksa}
+\cdag_{\ksa} \ketbra{a}{b}T_{ba}^*(\ksa)\right].
\label{EqHtunnel}
\end{equation}
The matrix element $T_{ba}(\ksa)$ is the scattering amplitude 
for an electron in the
state $\ksa$ tunneling from the lead onto the dot, thereby 
changing the dot state
from state $|a\rangle$ to a state $|b\rangle$.
Their evaluation is sketched in App.~\ref{AppCouplings}. Note that 
this amplitude vanishes
unless the number of electrons in state $|b\rangle$, $N_b$, equals
$N_a+1$. We will generally denote states such that the particle number
increases with the position in the alphabet of the denoting letter.

Before proceeding it is important to introduce a consistent notation
in order to
keep track of the many-particle states in the leads. 
A general state vector for the entire system is written as $|a g\rangle
=|a\rangle\otimes|g\rangle$, with
$|g\rangle=|\{N_{k\ell\sigma}\}\rangle$ 
denoting the state of both leads where $N_{\ksa}\in \{0,1\}$.
Throughout the
derivation of the general equations we use the  following notation to ensure
the anti-commutator rules of the operators
\begin{itemize}
\item $\ket{g-\ksa}\equiv \cno_{\ksa}\ket{g}$ and $\ket{g+\ksa}\equiv \cdag_{\ksa}\ket{g}$.\\
I.e. $\ket{g-\ksa}$ denotes the same set of indices as the state
$\ket{g}$,  but with $N_{\ksa}$ reduced by one. Furthermore it
contains a minus sign depending on the number of occupied states to the
left of the position $\ksa$. 
\item $\ket{g\ksa}\equiv
\cdag_{\ksa}\cno_{\ksa}\ket{g}$ and $\ket{g\overline{\ksa}~}\equiv
\cno_{\ksa}\cdag_{\ksa}\ket{g}$. I.e., 
$\ket{g\ksa}=\delta_{N_{\ksa},1}|g\rangle$.
\item The order of indices is
opposite to the order of the operators. E.g. $\ket{g-k'\sigma'
\ell'+\ksa}=\cdag_{\ksa}\cno_{k'\sigma'\ell'}\ket{g}
=-\cno_{k'\sigma'\ell'}\cdag_{\ksa}\ket{g}
=-\ket{g+\ksa-k'\sigma'\ell'}$ for $k\sigma\ell\neq k'\sigma'\ell'$, 
which is tacitly assumed, unless stated otherwise.
\end{itemize}
To simplify the notation, $\sigma\ell$ is only attached to $k$ 
the first time the index $k$ appears in the equation, 
and in the following it is implicitly assumed to be connected with $k$.
We also use the convention that $\sum_{k\sigma(\ell)}$ 
means summing over $k$ and
$\sigma$ with a fixed $\ell$, which is being connected to $k$ in
this sum.

The  matrix elements of the density operator $\hat{\rho}$ are
denoted $\dens{ag;bg'}=\bra{ag}\hat{\rho}\ket{bg'}$
and the time evolution of the matrix elements are governed by the
von Neumann equation
\begin{equation}\eqlab{neumann}
\imai\hbar\frac{\d}{\d
t}\dens{ag;bg'}=\bra{ag}H\hat{\rho}-\hat{\rho}H\ket{bg'}
\end{equation}

The particle current from the left lead into the structure, $J_L$, 
equals the rate of change in the occupation of the left lead. 
We find that
\begin{equation}
\begin{split}
J_L  & = -\frac{\d}{\d t}\sum_{k\sigma(L)}\ave{\cdag_{k}\cno_{k}}=
-\frac{\d}{\d t}\sum_{k\sigma(L)}\rho_{bg,bgk}\\
&=-\frac{2}{\hbar}\sum_{k\sigma(L),cb}
\Im\left\{\sum_{g}T^*_{cb}(k)\dens{cg-k;bg}\right\}
\label{EqCurrent}
\end{split}
\end{equation}
where we have used the definition of the density operator to 
calculate the average
value of the number operator in the left lead.

The goal is to determine these
elements of the density matrix, which describe the  correlations 
between the leads
and the dot. They are determined using the equation-of-motion
technique, and from \eqreff{neumann} we obtain
\begin{multline}
\eqlab{start}
\imai\hbar\frac{\d}{\d t} \dens{cg-\ksa;bg}=(E_c-E_b-E_k)\dens{cg-k;bg}\\
+\sum_{b'}T_{cb'}(k)\dens{b'gk;bg}-\sum_{c'}\dens{cg-k;c'g-k}T_{c'b}(k)\\
+\sum_{k'\sigma'\ell'}\Big[
\sum_{b'}T_{cb'}(k')\dens{b'g-k+k';bg}
+\sum_{d}T_{dc}^*(k')\dens{dg-k-k';bg}\\
-\sum_{c'}\dens{cg-k;c'g-k'}T_{c'b}(k')-\sum_{a}\dens{cg-k;ag+k'}T_{ba}^*(k')\Big].
\end{multline}
While $\dens{cg-\ksa;bg}$ describes the transition of an electron with
quantum number $k$ and spin $\sigma$ from lead $\ell$ to the
central region, terms like $\dens{b'g-k+k';bg}$ describe the
correlated transition of two electrons with $k$ and $k'$.
 $\dens{b'g-k+k';bg}$ satisfies a similar equation of motion
containing also correlated transition of three electrons on the right
hand side. 
In order to break the hierarchy we apply three approximations:

(i) We only consider coherent processes involving transitions of at 
most two different $k$-states. (ii) The time dependence of terms 
generating two-electron
transition processes is neglected, which corresponds to the Markov
limit. \cite{KuhnBookArt1998}
(iii) We assume that the level occupations $f_{k\sigma\ell}$ in the
leads are unaffected by
the kinetics of the dot, so it is possible to factorize the density
in the leads
and on the dot. This is realistic for "large" leads which are strongly
coupled to reservoirs, i.e. good contacts.

Defining
\begin{eqnarray}
w_{b'b}^{{}}&=&\sum_g\dens{b'g;bg},\label{EqDefW}\\
\phi_{ba}(\ksa) &=&\sum_g\dens{bg-k;ag}\label{EqDefPhi}
\end{eqnarray}
we find the following 
set of coupled differential equations (see Appendix \ref{AppDeriv}
for a detailed derivation): 
\begin{widetext}
\begin{equation}\begin{split}
\label{EqPhiDGL} 
\imai \hbar \frac{\d }{\d t}\phi_{cb}(k\sigma \ell)=&
(E_c-E_b-E_{k})\phi_{cb}(k)
+\sum_{b'}T_{cb'}(k)f_kw_{b'b}-\sum_{c'}(1-f_k)w_{cc'}T_{c'b}(k)
\\
+&\sum_{a,b',k'\sigma' \ell'} \frac{\left[T_{cb'}(k')f_{k'}\phi_{b'a}(k)
-(1-f_k)\phi_{cb'}(k')T_{b'a}(k)\right]T_{ba}^*(k')} {E_{k}+E_{k'}-(E_c-E_a)+\imai
0^+}
\\
+&\sum_{a,b',k'\sigma' \ell'} \frac{\left[(1-f_{k'})\phi_{cb'}(k)T_{b'a}(k')
-T_{cb'}(k)f_{k}\phi_{b'a}(k')\right]T_{ba}^*(k')} {E_{k}+E_{k'}-(E_c-E_a)+\imai
0^+}
\\
+&\sum_{a,b',k'\sigma' \ell'}
\frac{T_{cb'}(k')\left[f_{k'}\phi_{b'a}(k)T_{ba}^*(k')
-T_{b'a}(k)f_k\phi_{ba}^*(k')\right]} {E_{k}-E_{k'}-(E_{b'}-E_b)+\imai 0^+}
\\
+&\sum_{b',c',k'\sigma' \ell'}
\frac{T_{cb'}(k')\left[T^*_{c'b'}(k')(1-f_{k'})\phi_{c'b}(k)
-(1-f_k)\phi^*_{c'b'}(k')T_{c'b}(k)\right]} {E_{k}-E_{k'}-(E_{b'}-E_b)+\imai 0^+}
\\
+&\sum_{b',c',k'\sigma' \ell'} \frac{\left[f_{k'}\phi_{cb'}(k)T_{c'b'}^*(k')
-T_{cb'}(k)f_k\phi_{c'b'}^*(k')\right]T_{c'b}(k')} {E_{k}-E_{k'}-(E_c-E_{c'})+\imai
0^+}
\\
+&\sum_{c',d,k'\sigma' \ell'} \frac{\left[T_{dc}^*(k')(1-f_{k'})\phi_{dc'}(k)
-(1-f_{k})\phi^*_{dc}(k')T_{dc'}(k)\right]T_{c'b}(k')}
{E_{k}-E_{k'}-(E_c-E_{c'})+\imai 0^+}
\\
+&\sum_{c',d,k'\sigma' \ell'}
\frac{T_{dc}^*(k')\left[(1-f_{k'})\phi_{dc'}(k)T_{c'b}(k')
-T_{dc'}(k)f_k\phi_{c'b}(k')\right]} {E_{k}+E_{k'}-(E_d-E_b)+\imai 0^+}
\\
+&\sum_{c',d,k'\sigma' \ell'}
\frac{T_{dc}^*(k')\left[T_{dc'}(k')f_{k'}\phi_{c'b}(k)
-(1-f_k)\phi_{dc'}(k')T_{c'b}(k)\right]} {E_{k}+E_{k'}-(E_d-E_b)+\imai 0^+}
\end{split}\end{equation}
\begin{equation}
\imai \hbar \frac{\d }{\d t}w_{bb'}=(E_b-E_{b'})w_{bb'}+ \sum_{a,k\sigma\ell}
\left(T_{ba}(k)\phi^*_{b'a}(k)-\phi_{ba}(k)T^*_{b'a}(k)\right)
+\sum_{c,k\sigma\ell}
\left(T_{cb}^*(k)\phi_{cb'}(k)-\phi^*_{cb}(k)T_{cb'}(k)\right) 
\label{EqWDGL}
\end{equation}

\end{widetext}
These equations are the main result of this paper. 
They satisfy current conservation, as shown in 
App.~\ref{App:CurrentConservation}.
The numerical
implementation of this approach is straightforward  and we will give examples 
in the following sections. 

If we entirely neglect the correlated two-particle transitions, only
the first line of Eq.~(\ref{EqPhiDGL}) remains. Applying the Markov limit
we obtain a set of equations analogously to 
Eqs.~(2a,b) of Ref.~\onlinecite{GurvitzPRB1998}. This shows that our
approximation scheme goes substantially beyond the rate equation scheme of
Gurvitz \cite{GurvitzPRB1998}, which only holds
in the high-bias limit.

\section{Single level without spin}\label{SecSingleLevel}
In order to demonstrate the formalism described in the previous
section we consider a single level without spin.
We show that this case can be solved 
analytically in the stationary state and that the exact
nonequilibrium Green function result is recovered. 

The possible dot states are the empty state $0$ with energy $E_0=0$
and the occupied state $1$ with energy $\Ed$. The coupling matrix elements between the leads
and the dot are $T_{10}(k\ell)=T_{\ell}(k)$, and the others equal zero.\\
Inserting this in Eq.~(\ref{EqPhiDGL}) with $c=1$ and $b=0$ gives
\begin{equation}\label{phi10}\begin{split}
\imai\hbar\frac{\d}{\d t}&\phi_{10}(k\ell)
=\left[\Ed-E_k+\Sigma(E_k)\right]\phi_{10}(k\ell)\\
&- T_\ell(k)\sum_{k'\ell'}\frac{T_{\ell'}(k')\phi^*_{10}(k'\ell')}
{E_k-E_k'+\imai0^+}+T_\ell(k)(f_k-w_{11}),
\end{split}\end{equation}
where the self-energy
\begin{equation}\Sigma(E)=\sum_{k\ell}\frac{|T_\ell(k)|^2}{E-E_{k}+\imai 0^+}
\end{equation}
 has been introduced, and we have used the normalization of the probability
$w_{00}+w_{11}=1$.

After multiplying Eq.~(\ref{phi10}) with $T_\ell^*(k)\delta(E-E_k)$ and summing
over all $k$-states (in a fixed lead $\ell$) we obtain
\begin{equation}\begin{split}
\imai\hbar&\frac{\d}{\d t}B^{\ell}_{10}(E) =\left[\Ed-E+\Sigma(E)\right]B^{\ell}_{10}(E)\\
&-\frac{\Gamma_{\ell}(E)}{2\pi}\int \d E'\frac{B_{10}^{L*}(E')+B_{10}^{R*}(E')}{E-E'+\imai
0^+}\\
&+\frac{\Gamma_{\ell}(E)}{2\pi}[f_{\ell}(E)-w_{11}]
\label{EqDGLSingleDot}
\end{split}\end{equation}
for the new variable
\begin{equation}
B^{\ell}_{10}(E)=\sum_{k}\delta(E-E_{k})
T_{\ell}^*(k)\phi_{10}(k\ell)
\end{equation}
where $\Gamma_{\ell}(E)=2\pi\sum_{k}\delta(E-E_{k})|T_{\ell}(k)|^2$.\\
Eq.~(\ref{EqWDGL}) becomes
\begin{equation}\label{w11}
\frac{\d}{\d t}w_{11}^{{}}=-\frac{2}{\hbar}
\int\d E\,
\Im\left\{B^L_{10}(E)+B^R_{10}(E)\right\}\, .
\end{equation}
Finally, the current formula Eq.~(\ref{EqCurrent}) yields
\begin{equation}
J_L=-\frac{2}{\hbar}\int \d E ~\Im\left\{B^L_{10}(E)\right\}\, .
\label{EqJSingle}
\end{equation}

Throughout this paper, we apply Fermi functions 
$f_{k\sigma\ell}=1/[\exp((E_k-\mu_{\ell})/k_BT)+1]\equiv 
f_{\ell}(E_k)$ for the lead
occupations with chemical potentials $\mu_{\ell}$ and temperature $T$.
Except for this section, the bias $V$ is applied symmetrically around zero,
i.e. $\mu_L=V/2$, $\mu_R=-V/2$.
The contact functions
$\Gamma_{\ell}(E)$ are assumed to be zero for $|E|>W$, while they take
the constant values $\Gamma_{\ell}$, independent of spin, for 
$|E|<0.95W$. For $0.95W<|E|<W$ we interpolate with an elliptic
behavior in order to avoid discontinuities. 

\begin{figure}
\includegraphics[width=0.9\columnwidth,keepaspectratio]{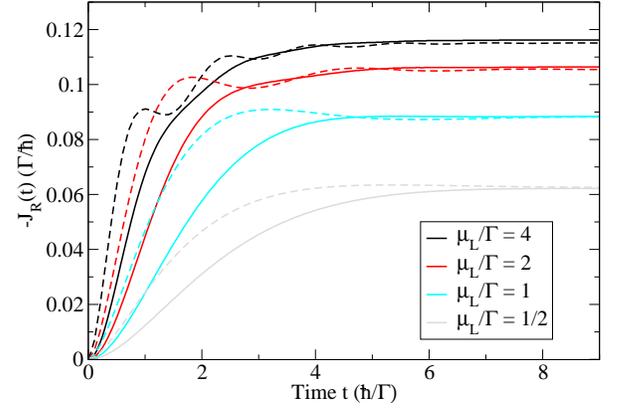}
\caption[a]{(Color online) The time-dependent current
calculated with our 
method (full line) and with the time-dependent Green function 
method from Ref.~\onlinecite{StefanucciPRB2004} (dashed line)
as response to a step-like modulation of the bias with step height $\mu_L$. 
The coupling is $\Gamma_L=\Gamma_R=\Gamma/2$, the temperature $k_BT=0.05\Gamma$,
and the half-width of the band is $W=30\Gamma$.}
\label{FigStefanucci}
\end{figure}

The time-dependent net-current $J_R(t)$ flowing from the right lead
into the single level has been calculated from
Eqs.~(\ref{EqDGLSingleDot},\ref{w11},\ref{EqJSingle}) in the 
following situation: For times $t<0$
the chemical potentials of both leads and the single level are aligned, i.e.
$\mu_L^0=\mu_R^0=E_1=0$. At $t=0$ the chemical potential of the
left lead is raised instantaneously to $\mu_L$ giving a
step-like modulation of the bias. 
The result is shown for different values of $\mu_L$ in 
Fig.~\ref{FigStefanucci}.
Also shown is the result of an exact time-dependent Green function
calculation.\cite{StefanucciPRB2004} 
It is not surprising that our 
results do not show the exact time-dependence because the Markov
limit has been invoked in the derivation of the generalized
equation system in Eqs.~(\ref{EqPhiDGL},\ref{EqWDGL}).

In the long-time limit, we reach a stationary state with the current
\begin{equation}\label{EqJSingleStat}
J_L=J_R=\frac{1}{\hbar}\int \frac{\d
E}{2\pi}\frac{\Gamma_L(E)\Gamma_R(E)[f_L(E)-f_R(E)]}{|E-\Ed-\Sigma(E)|^2},
\end{equation}
which is derived analytically in App.~\ref{AppDerJSingle}. Eq.~(\ref{EqJSingleStat})
is in full agreement with the exact nonequilibrium Green function
result. \cite{MeirPRL1992} 

\section{Double quantum dot}\label{SecDoubleDot}
The double quantum dot structure, where the dots are coupled in
series, is a standard example to study
tunneling through a multiple-level system. In case of Coulomb
interaction and finite bias the validity of both the rate equation 
method and the Green function formalism is limited.

To simplify the
analysis we treat the spinless case. (A possible realization is to favor one
spin polarization of the electron by a high magnetic field.)
Denoting the left/right dot by
$\alpha/\beta$, the Hamiltonian reads
\begin{equation}
\begin{split}
H=&E_\alpha d_\alpha^\dag d_\alpha^{{}} +E_\beta d_\beta^\dag d_\beta^{{}}+U d_\alpha^\dag d_\alpha^{{}} d_\beta^\dag d_\beta^{{}}\\
&+\left(\Omega d_\beta^\dag d_\alpha^{{}}
+h.c\right)
+\sum_{k\ell}E_{k\ell}^{{}} c^\dagger_{k\ell} c^{{}}_{k\ell}\\
&+\sum_{k}\Big( t_{kL}^{{}}d_\alpha^\dag c_{kL}^{{}}
+t_{kR}^{{}}d_\beta^{{\dag}}c_{kR}^{{}}+h.c.\Big)
\end{split}
\end{equation}
with $\Omega$ being the interdot tunneling coupling and $U$ the
Coulomb energy for occupying both dots. The first four terms
describe the isolated double quantum dot $H_D$.
Diagonalizing this part of the Hamiltonian gives the following
states
\begin{alignat*}{2}
\ket{0} &= \ket{0},&  E_0&=0,\\
\ket{1} &=(\alpha_1d_\alpha^{\dag}+\beta_1d_\beta^{\dag})\ket{0}, & \quad
E_1 &=\frac{1}{2}\left[\rho-\sqrt{\Delta^2+4\Omega^2}\right],\\
\ket{2} &=(\alpha_2d_\alpha^{\dag}+\beta_2d_\beta^{\dag})\ket{0}, & \quad
E_2 &=\frac{1}{2}\left[\rho+\sqrt{\Delta^2+4\Omega^2}\right],\\
\ket{d} &=d_\alpha^{\dag}d_\beta^{\dag}\ket{0},& E_d&=  E_\alpha+E_\beta+U,
\end{alignat*}
with $\Delta=E_\alpha-E_\beta$, $\rho=E_\alpha+E_\beta$,
$C_{\pm}=\frac{\Delta\pm\sqrt{\Delta^2+4\Omega^2}}{2\Omega}$,
$\alpha_{1/2}=\frac{C_\mp}{\sqrt{1+C^2_{\mp}}}$ and
$\beta_{1/2}=\frac{1}{\sqrt{1+C^2_{\mp}}}$.

From Appendix~\ref{AppCouplings}  we find
\begin{equation}
\begin{split}
H_T=&\sum_{k\ell}\Big[
T_{10}^{\ell}(k)\ket{1}\bra{0}c_{k{\ell}}
+T_{20}^{\ell}(k)\ket{2}\bra{0}c_{k{\ell}}\\
&+T_{d1}^{\ell}(k)\ket{d}\bra{1}c_{k{\ell}}
+T_{d2}^{\ell}(k)\ket{d}\bra{2}c_{k{\ell}}\Big] +h.c.
\end{split}
\end{equation}
with (skipping the k-dependence of the matrix elements)
\begin{alignat*}{2}
T_{10}^L&=t_L^*\bra{1}d_\alpha^{\dag}\ket{0}=t_L^*\alpha_1^*, &
T_{20}^L&=t_L^*\bra{2}d_\alpha^{\dag}\ket{0}=t_L^*\alpha_2^*,\\
T_{10}^R&=t_R^*\bra{1}d_\beta^{\dag}\ket{0}=t_R^*\beta_1^*, &
T_{20}^R&=t_R^*\bra{2}d_\beta^{\dag}\ket{0}=t_R^*\beta_2^*,\\
T_{d1}^L&=t_L^*\bra{d}d_\alpha^{\dag}\ket{1}
=t_L^*\beta_1, & T_{d2}^L&=t_L^*\bra{d}d_\alpha^{\dag}\ket{2}
=t_L^*\beta_2,\\
T_{d1}^R&=t_R^*\bra{d}d_\beta^{\dag}\ket{1}
=-t_R^*\alpha_1,&\quad
T_{d2}^R&=t_R^*\bra{d}d_\beta^{\dag}\ket{2}
=-t_R^*\alpha_2,
\end{alignat*}
where the signs of the coupling matrix elements are due to the
order of the operators in the double-occupied state.

Applying the method in the same way as for the single level system
gives eight different functions of the type $B^\ell_{cb}(E)$ and
five different occupations $\omega_{bb'}$. These equations have
been solved and the stationary current has been recorded.

\begin{figure}
\includegraphics[width=0.9\columnwidth,keepaspectratio]{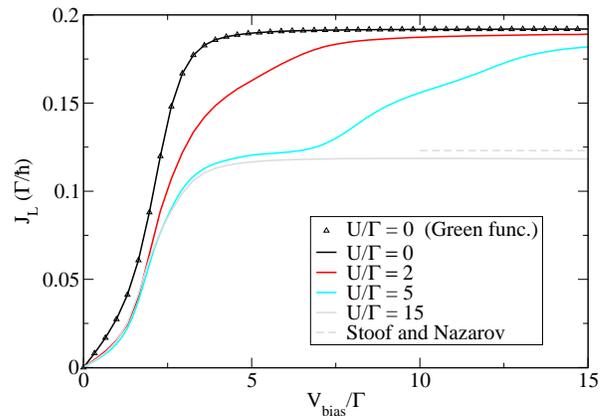}
\caption[a]{(Color online) Stationary current through the double
quantum dot structure for different values of the  interdot Coulomb
repulsion $U$. The triangles are from a nonequilibrium Green
function calculation, and the dotted line is the result by Stoof and
Nazarov  \cite{StoofPRB1996} valid in high-bias limit for
$U\rightarrow\infty$. The levels of the dot are placed symmetrically
around the zero-bias with $E_\alpha-E_\beta=\Gamma$. We use
the interdot tunneling coupling $\Omega=\Gamma$,
$\Gamma_L=\Gamma_R=\Gamma/2$, the temperature $k_BT=0.1\Gamma$, and
the half-width of the band $W=20\Gamma$.}
\label{FigDQD}
\end{figure}

By comparing with exact Green function results, it has been
verified numerically  that in the non-interacting case ($U=0$) the
exact transmission is obtained for various values of level
splitting and interdot coupling (not shown). Furthermore, for both
$U=0$ and non-zero $U$ we have calculated the stationary current
in a situation where the levels are de-aligned with
$E_\alpha=-E_\beta=0.5\Gamma$, and $\Omega=\Gamma$. 
The results for different values of $U$ are shown in Fig.~\ref{FigDQD} 
together with the Green function result for $U=0$. Obviously, the
latter is fully recovered in the non-interacting limit.
The straight dashed line in the figure is the quantum rate equation
result obtained by Stoof and Nazarov\cite{StoofPRB1996}, which is valid
in the high-bias limit ($V\to \infty$) for $U\rightarrow\infty$. The same result
is found in Ref.~\onlinecite{GurvitzPRB1996}
using another rate equation method. 
The small discrepancy between the results could
be due to the finite  bandwidth used in our calculation.
For intermediate values of $U$ the results looks reasonable
and exhibit a smooth interpolation between the limiting cases. The
kink on the curve for finite $U$ is due to the single occupied
state.

\section{Spin-degenerate level}\label{SecDoubleSpin}
\begin{figure}
\includegraphics[width=0.9\columnwidth,keepaspectratio]{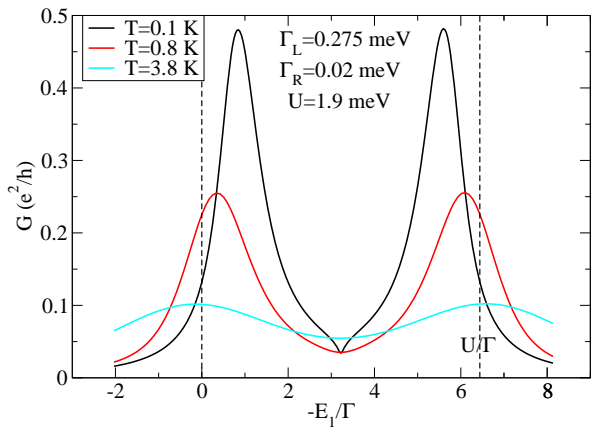}
\caption[a]{(Color online) Zero-bias conductance as a function of 
level position
for different temperatures. All parameters are according to the
experimental data shown in Fig.~2 of Ref.~\onlinecite{GoldhaberPRL1998}.}
\label{FigGoldhaberf2}
\end{figure}

Now we consider a spin-degenerate single level with energy $E_1$ and
Coulomb interaction $U$. 
We use the parameters $U=1.9$ meV,
and $\Gamma=\Gamma_L+\Gamma_R=0.295$ meV, as 
experimentally determined for the structure studied in 
Ref.~\onlinecite{GoldhaberPRL1998}.
The conductance 
\begin{equation}
G=e^2\frac{\d J}{\d(\mu_L-\mu_R)}
\end{equation}
is expected to reach 
$G_0=\frac{e^2}{h}\,8\Gamma_L\Gamma_R/(\Gamma_L+\Gamma_R)^2$
in the  zero-bias limit $\mu_L\to \mu_R$
for temperatures far below the Kondo temperature $T_K$ 
\cite{NgPRL1988,GlazmanJETP1988}.
As $G_{\rm max}\approx 0.5 e^2/h$ in the experiment we use 
$\Gamma_L=0.275$ meV and $\Gamma_R=0.02$ meV. 
Furthermore the band width $W=5$ meV is applied.
In Fig.~\ref{FigGoldhaberf2} we show the zero-bias conductance
as a function of the dot level, which is modified by a gate bias in
the experiment. We find the standard Coulomb oscillations, 
where the conductance exhibits peaks whenever 
the single-particle excitation energies are close to the Fermi edge of
the contacts, $\mu=0$
(depicted by vertical dashed lines at $E_1=0$ and $E_1=-U$).
The peak positions and widths are in good agreement
with the data given in Fig.~2 of Ref.~\onlinecite{GoldhaberPRL1998}. 
The peak heights for the  peak around
$E_1\approx 0$ agree reasonably with the experiment, if one takes 
into account that for elevated  temperatures the presence of
different levels raise the conductance which is not included 
in our single-level model. (The experimental level spacing 
corresponds to 5 K.)
The experimental peak heights for the peak at $E_1\approx -U$
are lower, while they are exactly identical with the corresponding
peaks $E_1\approx 0$ due to  electron-hole symmetry in our
calculation. Possible sources for this deviation result from  
an energy-dependence
of the $\Gamma_{\ell}(E)$ in the experiment or the admixture of different
levels.

Further lowering the temperature, the zero-bias 
conductance should increase in the region $0<E_1<U$, due to the Kondo
effect \cite{NgPRL1988,GlazmanJETP1988}. Albeit we observe an increase
in parts of this region, the (probable unphysical) dip in our curve
for $T=0.1$ K at $E_1=-U/2$ persists even at lower temperatures.
Furthermore the conductance can exceed $G_0$ at the
peaks. This indicates that our approach fails in the Kondo limit, where
strong correlations between lead and dot state require
elaborated renormalization group
\cite{CostiJPC1994,SchoellerPRL2000,RoschPRL2003} 
or slave boson \cite{WingreenPRB1994,DongJPC2001} techniques.

\begin{figure}
\includegraphics[width=0.95\columnwidth,keepaspectratio]{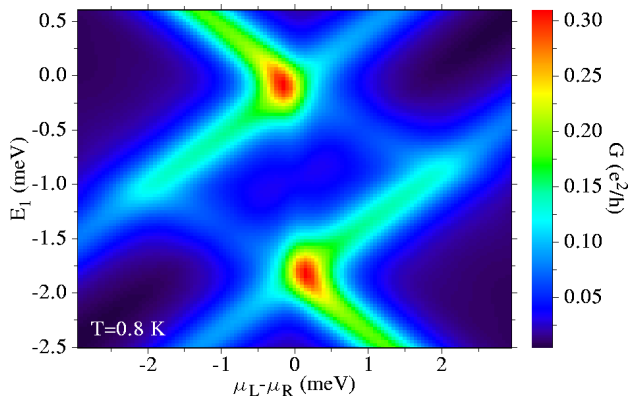}
\caption[a]{(Color online) Differential conductance for finite bias. 
Parameters as in Fig.~\ref{FigGoldhaberf2}.}
\label{FigKondoCondArea}
\end{figure}

\begin{figure}
\includegraphics[width=0.9\columnwidth,keepaspectratio]{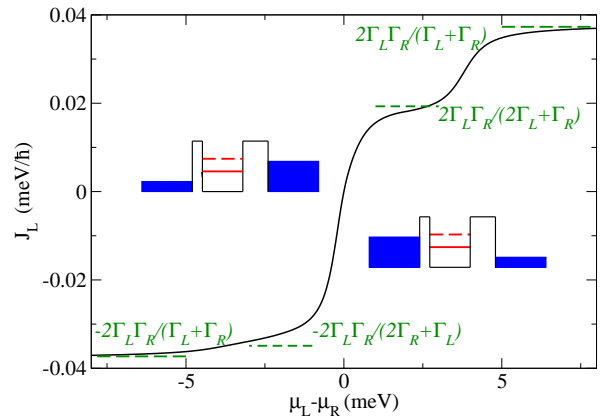}
\caption[a]{(Color online) Current versus bias for $E_1=0$ and $T=0.8$ K.
Parameters as in Fig.~\ref{FigGoldhaberf2}. The dashed  horizontal 
lines correspond to the rate equation model by Gurvitz and 
Prager\cite{GurvitzPRB1996},
where we added the respective formulae.}
\label{FigcurrentDotSpin}
\end{figure}

In Fig.~\ref{FigKondoCondArea} we show the finite bias 
conductance at 0.8 K, where both the conductance peaks for
$\mu_L\approx \mu_R$ discussed
above as well as the excitations can be detected.
We observe a strong asymmetry due to $\Gamma_L\gg\Gamma_R$.
This can be understood from Fig.~\ref{FigcurrentDotSpin}, where
the current is plotted versus bias at the single-electron excitation peak
$E_1=0$.
For negative bias the electrons rapidly leave the
dot via the thin left barrier and the dot is essentially empty.
Thus both spin directions can 
tunnel through the thick right barrier, which is limiting the current.
In contrast, for positive bias the dot is occupied with a single
electron (as long as $\mu_L=V/2<E_1+U$) with a given spin
and only this spin direction may tunnel through the thick
barrier, reducing the current approximately by a factor of 2. 
We have shown the respective results for the rate equation model
\cite{GurvitzPRB1996} for comparison. 
The short-dashed horizontal lines refer to 
a bias which allows only single occupation of the dot, while
the long-dashed line considers the case where both the single- and
the two-particle state are located between both Fermi levels.
The currents from the rate equation model slightly exceed our results, as
the peaks are not completely within the bias window due to
broadening.

\section{Discussion and summary} 

We have presented an approach for transport through finite systems
based on the Liouville equation. This approach recovers the results
from the Green-function method in the noninteracting limit for the
models studied. In the high-bias limit the results are consistent
with the  many-particle rate equations.
Thus it bridges
the gap between these approaches and allows for a consistent treatment
of Coulomb interaction and broadening effects for arbitrary bias.
E.g. Coulomb blockade peaks are correctly reproduced.
The model fails below the Kondo temperature where strong
correlations between the finite system and the 
contacts dominate the behavior.

Correlations between tunneling events
have been previously studied by the method of a
diagrammatic real-time technique \cite{KonigPRL1997}. While this work
was completed we also became aware of a cumulant expansion of the 
tunneling Hamiltonian \cite{LiPRB2005}. It would be interesting to study the
relation between these approaches and our method. A central question
is here, whether the exact Green function result, 
such as Eq.~(\ref{EqJSingleStat}), can be obtained for the
noninteracting case.

The numerical implementation of our approach
is straightforward and explicit results were presented for standard
model systems made by up to two single-particle states. For larger 
systems the number of many-particle states $b,c$
increases dramatically, and so does the  number of
$\phi_{bc}$-functions. Thus sophisticated routines are
needed for the implementation and evaluation of real systems.

\acknowledgments
The authors thank J. Paaske for helpful discussions regarding the Kondo effect.
This work was supported by the Swedish Research Council (VR).

\appendix
\section{Determination of matrix elements $T_{ab}(k)$\label{AppCouplings}}

Conventually one starts with  a single-particle basis  in the central
region with wave functions 
$\Psi_n({\bf r})$, spin functions $\chi_{\sigma}$ and 
associated creation operators $d^{\dag}_{n\sigma}$.
Then an arbitrary many-particle state $|a\rangle$
can be written as
\[
|a\rangle=\sum_{\bf j}a_{\bf j}
d^{\dag}_{j_{1}}
d^{\dag}_{j_{2}}
\ldots d^{\dag}_{j_{N_a}}
|0\rangle
\]
where $j_i=n_i\sigma_i$ determines the $i$-th single-particle state
in the $N_a$-particle Slater determinant determined by the index set
${\bf j}=(j_1,j_2,\ldots j_{N_a})$. In order to avoid double counting,
we restrict to the ordering $n_1\le n_2\le \ldots \le n_{N_a}$, where
spin-up precedes spin-down for equal $n$.  
The expansion
coefficients $a_{\bf j}$ can be obtained by exact diagonalization
of the dot Hamiltonian.

In the single-particle basis the tunneling Hamiltonian reads
\begin{equation}
H_T=
\sum_{k\sigma\ell,n}
\left(
t^*_{n\ell}(k\sigma\ell)c^{\dag}_{k\sigma\ell}d_{n\sigma}
+t_{n\ell}(k\sigma\ell)d_{n\sigma}^{\dag}c_{k\sigma\ell}
\right)
\end{equation}
Inserting the unit operators $\sum_a |a\rangle\langle a|$,
$\sum_b |b\rangle\langle b|$
we find
\begin{equation}\begin{split}
H_T=
\sum_{k\sigma\ell,a,b}&\Big(
c^{\dag}_{k\sigma\ell}|b\rangle 
\underbrace{\sum_{n}t^*_{n\ell}(k\sigma\ell)
  \langle b|d_{n\sigma}|a\rangle}_{=T^{*}_{ab}(k\sigma\ell)}\langle a|\\
 &+|a\rangle \underbrace{\sum_{n}t_{n\ell}(k\sigma\ell)
\langle a|d^{\dag}_{n\sigma}|b\rangle}_{=T_{ab}(k\sigma\ell)}\langle b|
c_{k\sigma\ell}
\Big)
\end{split}\end{equation}
to be used in Eq.~(\ref{EqHtunnel}).

\section{Derivation of Eqs.~(\ref{EqPhiDGL},\ref{EqWDGL})
\label{AppDeriv}}

Using the approximation (i) we find for one of
the two-electron transition terms in \eqreff{start}
\begin{multline*}
\imai\hbar\frac{\d}{\d t} \dens{b'g-\ksa+k'\sigma'\ell';bg}
=(E_{b'}+E_{k'}-E_b-E_k)\dens{b'g-k+k';bg}\\
-\sum_a T_{b'a}(k)\dens{agk+k';bg}
+\sum_{c'} T_{c'b'}^*(k')\dens{c'g-k\overline{k'};bg}\\
-\sum_{c'}\dens{b'g-k+k';c'g-k}T_{c'b}(k)
-\sum_a\dens{b'g-k+k';ag+k'}T_{ba}^*(k')
\end{multline*}
Now we take the Markov limit (ii) following the standard treatment of
density matrix theory for ultrafast dynamics \cite{KuhnBookArt1998}.
This implies adding $-\iop\dens{b'g-\ksa+k'\sigma'\ell';bg}$
on the right hand side in order to guaranty the decay of
initial conditions at $t=-\infty$, and
neglecting the time dependence of the inhomogeneity\footnote{This
becomes exact for the stationary state, but can, however, produce 
incorrect results for the time-dependence.} 
(second and third
line). Then 
this linear differential equation can be solved directly, yielding 
\begin{multline*}
\dens{b'g-\ksa+k'\sigma'\ell';bg}=
\frac{1}{E_k-E_{k'}-(E_{b'}-E_b)+\iop}\\
\times\Big[-\sum_a T_{b'a}(k)\dens{agk+k';bg}
+\sum_{c'} T_{c'b'}^*(k')\dens{c'g-k\overline{k'};bg}\\
-\sum_{c'}\dens{b'g-k+k';c'g-k}T_{c'b}(k)
-\sum_a\dens{b'g-k+k';ag+k'}T_{ba}^*(k')\Big]
\end{multline*}
In the same way the other two-electron transition terms in \eqreff{start}
are determined by
\begin{multline*}
\dens{dg-\ksa-k'\sigma'\ell';bg}=
\frac{1}{E_k+E_{k'}-(E_d-E_b)+\iop}\\
\times\sum_{c'}\Big[-T_{dc'}(k)\dens{c'gk-k';bg}+T_{dc'}(k')\dens{c'g-kk';bg}\\
-\dens{dg-k-k';c'g-k}T_{c'b}(k)-\dens{dg-k-k';c'g-k'}T_{c'b}(k')\Big],
\end{multline*}

\begin{multline*}
\dens{cg-\ksa;c'g-k'\sigma'\ell'}=
\frac{1}{E_k-E_{k'}-(E_c-E_{c'})+\iop}\\
\times\Big[\sum_{b'}T_{cb'}(k)\dens{b'gk;c'g-k'}+\sum_{d}T_{dc}^*(k')\dens{dg-k-k';c'g-k'}\\
-\sum_{d}\dens{cg-k;dg-k'-k}T_{dc'}(k)-\sum_{b'}\dens{cg-k;b'gk'}T_{c'b'}^*(k')\Big],
\end{multline*}

\begin{multline*}
\dens{cg-\ksa;ag+k'\sigma'\ell'}=
\frac{1}{E_k+E_{k'}-(E_c-E_a)+\iop}\\
\times\sum_{b'}\Big[T_{cb'}(k)\dens{b'gk;ag+k'}+T_{cb'}(k')\dens{b'g-k+k';ag+k'}\\
-\dens{cg-k;b'g+k'-k}T_{b'a}(k)-\dens{cg-k;b'g\overline{k'}}T_{b'a}(k')\Big].
\end{multline*}

In order to obtain Eq.~(\ref{EqPhiDGL})
we sum over $g$ in \eqreff{start} after inserting the above
approximations for the two-electron transition terms. Using 
the definitions (\ref{EqDefW},\ref{EqDefPhi}) and 
the decoupling assumption (iii) we obtain
\begin{equation*}
\sum_g
\rho_{b'gk;bg}=\sum_g\delta_{N_{k},1}\rho_{b'g;bg}
\approx f_k\sum_g \rho_{b'g;bg}=f_kw_{b'b}
\end{equation*}
Similarly
\begin{equation*}
\begin{split}
\sum_g\rho_{b'g\overline{k};bg}&\approx (1-f_k)w_{b'b},\\
\sum_g\rho_{bg-k'k ;ag}&\approx  f_k\phi_{ba}(k'),\\
\sum_g\rho_{bg-k'\overline{k};ag}&\approx  (1-f_k)\phi_{ba}(k')
\end{split}
\end{equation*}
Furthermore note that 
\[\begin{split}
\sum_g\rho_{bg;ag+k}=&\sum_{\tilde{g}}\rho_{b\tilde{g}-k;a\tilde{g}}=\phi_{ba}(k)\\
\sum_g\rho_{b'g+k;bg+k}=&\sum_{\tilde{g}}\rho_{b'\tilde{g}k;b\tilde{g}}\approx
f_kw_{b'b}
\end{split}\]
as well as similar relations hold, where $|\tilde{g}\rangle$ is identical with $|g\rangle$
except for exchanging 1 and 0 in the occupation of state $k$
(including the appropriate change of sign).
Particular care has to be taken in order to insure the anti-commutation
rules. E.g., $\sum_g\dens{b'g-k+k';ag+k'}\approx
-f_{k'}\phi_{b'a}(k)$.

In the same way 
\begin{equation*}
\begin {split}
\imai\hbar&\frac{\d}{\d t} \dens{bg;b'g}=(E_b-E_{b'})\dens{bg;b'g}\\
&+\sum_{a,\ksa}T_{ba}(k)\dens{ag+k;b'g}
+\sum_{c,\ksa}T_{cb}^*(k)\dens{cg-k;b'g}\\
&-\sum_{c,\ksa}\dens{bg;cg-k}T_{cb'}(k)
-\sum_{a,\ksa}\dens{bg;ag+k}T_{b'a}^*(k).
\end{split}
\end{equation*}
gives Eq.~(\ref{EqWDGL}) after summing over $g$.

\section{Conservation of current}\label{App:CurrentConservation}
We will in the following show that the formalism obeys current conservation, i.e. 
\begin{equation}
\label{Eqcurrcons}
\frac{\d}{\d t}\ave{\hat{N}_D}=J_L+J_R,
\end{equation}
with $\hat{N}_D=\sum_b N_b \ketbra{b}{b}$ being the number operator of the dot.
From the definition of the density operator we get
\begin{equation}\eqlab{dotcons}
\begin{split}
\frac{\d}{\d t}\ave{\hat{N}_D}&=\frac{\d}{\d t}\text{Tr}
\left[\rho \hat{N}_D\right]
=\sum_{b}N_b\frac{\d}{\d t} w_{bb}.
\end{split}
\end{equation}
The time derivative of $w_{bb}$ is obtained from Eq.~(\ref{EqWDGL})
\begin{equation}
\begin{split}
\frac{\d}{\d t}w_{bb}&=-\frac{2}{\hbar}\sum_{a\ksa}\Im\left\{T_{ba}^*(k)\phi_{ba}(k)\right\}\\
&\quad+\frac{2}{\hbar}\sum_{c\ksa}\Im\left\{T_{cb}^*(k)\phi_{cb}(k)\right\}.
\end{split}
\end{equation}
Inserting this in \eqreff{dotcons} and renaming the summation
indices in the second term leads to
\begin{equation}\eqlab{dotN}
\frac{\d}{\d t}\ave{\hat{N}_D}=
-\frac{2}{\hbar}\sum_{ba\ksa}(N_b-N_a)\Im\left\{T_{ba}^*(k)\phi_{ba}(k)\right\}.
\end{equation}
Now the $T_{ba}(k)$-matrix elements are vanishing for
$N_b\neq N_a+1$, and the right-hand side of \eqreff{dotN} becomes
$J_L+J_R$ using the definition of the currents  Eq.~(\ref{EqCurrent}).
Thus, current conservation (\ref{Eqcurrcons}) holds.

\section{Derivation of Eq.~(\ref{EqJSingleStat})\label{AppDerJSingle}}

Defining $B_{10}=B^L_{10}+B^R_{10}$ and $\Gamma=\Gamma^L+\Gamma^R$ we
find from  Eq.~(\ref{EqDGLSingleDot})
\begin{multline}\label{EqDiffB}
\imai\hbar\frac{\d}{\d t}B_{10}(E)
=\left[\Ed-E+\Re\left\lbrace\Sigma(E)\right\rbrace\right]B_{10}(E)\\
+ \Gamma(E)\Im\left\lbrace B_{10}(E)\right\rbrace 
-\frac{\Gamma(E)}{2\pi}\int \d E'~{\cal P}\left\lbrace
\frac{B^*_{10}(E')}{E-E'}\right\rbrace \\
+ \frac{\Gamma_L(E)f_L(E)+\Gamma_R(E)f_R(E)}{2\pi}-
w_{11}\frac{\Gamma(E)}{2\pi}
\end{multline}
where $\Im \left\lbrace \Sigma(E)\right\rbrace =-\Gamma(E)/2$ has been used.
Eq.~(\ref{EqDiffB}) is a
linear inhomogeneous differential equation which has a particular 
stationary real solution $B^{{\rm stat}}_{10}(E)$ determined by
\begin{multline}
\eqlab{EqBStat}
\frac{\Gamma(E)}{2\pi}\int \d E'~{\cal P}\left\lbrace 
\frac{B^{\text{stat}}_{10}(E')}{E-E'}\right\rbrace\\
=\left[\Ed-E+\Re\left\lbrace\Sigma(E)\right\rbrace\right]
B^{\rm stat}_{10}(E)\\
+ \frac{\Gamma_L(E)f_L(E)+\Gamma_R(E)f_R(E)}{2\pi}-
w_{11}\frac{\Gamma(E)}{2\pi}
\end{multline}
Numerically, we find that this solution is indeed reached
from different initial conditions in the long-time limit.
Inserting the integral over $B^{\text{stat}}_{10}(E)$ from 
\eqreff{EqBStat} into Eq.~(\ref{EqDGLSingleDot}) gives the stationary solution
\begin{multline}
[\Ed-E+\Sigma(E)]B^L_{10}(E)\\
=\Gamma_L(E)\frac{\Ed-E+\Sigma(E)}{\Gamma(E)}B^{\rm  stat}_{10}(E) \\
+\frac{\Gamma_L(E)\Gamma_R(E)[f_R(E)-f_L(E)]}{2\pi\Gamma(E)}.
\end{multline}
As $B^{\text{stat}}_{10}(E)$ is real it does not contribute to
the imaginary part of $B^L_{10}(E)$ in Eq.~(\ref{EqJSingle})
providing the final result (\ref{EqJSingleStat}).


\end{document}